\documentclass{article}
\usepackage[utf8]{inputenc}
\usepackage{xcolor}
\usepackage{authblk}
\usepackage{url}

\title{Clouds in Exoplanetary Atmospheres} % Sample title
\author[1]{Christiane Helling}
\affil[1]{Centre for Exoplanet Science,  University of St Andrews, St Andrews, KY16 9SS, UK\newline
SUPA, School of Physics \& Astronomy, Univ. of St Andrews, St Andrews, KY16 9SS, UK\newline
SRON Netherlands Institute for Space Research, Sorbonnelaan 2, 3584 CA Utrecht, NL.}
\date{}
\usepackage{natbib}
\usepackage{graphicx}
\usepackage{geometry}
\usepackage[utf8]{inputenc}

\begin{document}
\maketitle
% Nominal suggested length of essay is 2000-3000 words. Slightly less or more is acceptable. 

\section{Introduction} % 400 words 
%\textcolor{red}{Brief overview of the area, placing it in the larger context of exoplanet science ($\sim$200 words). \newline 
%Big science questions/goals - bullet point list ($\sim$200 words)} 

Today, we know 4330 exoplanets orbiting their host stars in 3200
planetary systems\footnote{\url{http://exoplanet.eu}} The diversity of these exoplanets is huge, and none of
the known exoplanets is a twin to any of the solar system planets, nor
is any of the known extrasolar planetary systems a twin of the solar
system. Such diversity on many scales and structural levels requires
fundamental theoretical approaches.  Large efforts are underway to
develop fundamental theoretical approaches for individual aspects of
exoplanet sciences, like exoplanet atmospheres, cloud formation, disk
chemistry, planet system dynamics, mantle convection, mass loss of planetary atmospheres. 

The diversity of exoplanets calls for a fundamental, for an
 interdisciplinary and for an intelligent approach in order to fully
 understand their atmospheric composition.  Many harbour a dynamic
 atmosphere where cloud particles can form in abundance. Some planets
 will be fully covered in clouds, some have clouds on the nightside
 but are largely cloud-free on the dayside.  The basic processes of
 cloud formation are understood, and the main challenge lays in the
 fundamental physics and chemistry of material properties.  The
 simultaneous development of different cloud models (e.g., \cite{2006A&A...455..325H,2018ApJ...853....7K}) has enabled great
 insight into individual processes which now need to be combined. Such
 a combination requires an intelligent approach for tackling the
 micro-physical complexity involved. This includes the modelling of
 cluster formation, particle-particle growth/destruction, ionisation
 and photo-processes, diffusion and turbulent transport consistently
 coupled. Cloud models are a key part of the virtual laboratories that
 enable us to study the large parameter range of exoplanets, and to
 decipher the combination of individual observations.

Several of the known exoplanets orbit their host star very closely
such that a rocky surface turns into magma ocean (e.g. CoRot-7b or 55
Cnc e) or Jupiter-like exoplanets expand their atmosphere
considerably (e.g. HD\,189733b, HD\,209456b) and asymmetrically
(e.g. HAT-P-7b, WASP-18b). Inside the atmospheres of exoplanets, clouds form.
The cloud particles ({\it aerosols}) are made of a mix of minerals,
for example, silicates and iron oxide in giant gas planets \cite{2019AREPS..47..583H}, and sulfur hazes \cite{2017AJ....153..139G}. The 
global cloud composition changes with changing element abundance which can be affected by outgasing (e.g.for 55 Cnc e; \cite{2017MNRAS.472..447M}) and pre-determined by the location of planet formation within the disk \cite{2014Life....4..142H,2019A&A...632A..63C,2020A&A...633A.110G}. The composition of the upper crust of a rocky planet does determine the atmospheric element abundances, and eventually, the stability of liquid water \cite{Oli2019}.
Hydrocarbon hazes have been shown to form in the upper atmospheres of super-Earths and mini-Neptunes (e.g. GJ~1214b and GJ436b; \cite{2019ApJ...876L...5K}) under the effect of the UV radiation field of the M-dwarf type host stars. Similar effects have been proposed for the day-side of the giant gas planet WASP-43b where the UV photons of a K star enable the formation of hydrocarbon hazes in tandem with mineral cloud particles \cite{2020arXiv200514595H}. 
Spectroscopic studies of exoplanets are repeatedly frustrated by
clouds and/or hazes blocking the view into the underlying atmosphere to reveal the
existence of potential biosignatures, signature for their evolutionary state, or traces of planet formation. Clouds on Earth are linked to
the support of the terrestrial cycle of life as they can preserve the
right temperature of the biosphere and enable the transport of water
through the atmosphere. Cary exoplanet clouds as similar significance?

The following challenges need to be addressed in order to answer this
question, and, simultaneously provide the opportunity to progress our
understanding of exoplanets and their atmospheres by exploring our
models as virtual laboratories to fill gaps in observational data from
different instruments and missions, and taken at different instances
of times:\\[0.1cm] {\sf Challenge a)} Building {\it complex models
based on theoretical rigor} that aim to understand the interactions of
atmospheric processes, to treat cloud formation and its feedback onto
the gas-phase chemistry and the energy budget of the planetary
atmosphere moving away from solar-system inspired parameterisations.\\[0.1cm]
{\sf Challenge b)} Enabling {\it Cloud modelling based on fundamental
  physio-chemical insights} in order to be applicable to the large and
unexplored chemical, radiative and thermodynamical parameter range of
exoplanets in the universe.  Challenge b) will be explored in what
follows.

\section{State of the art - Cloud formation modelling}  % 500 words 
%\textcolor{red}{Brief summary of where we stand in this area. A short paragraph ($\sim$150 words) about origin of the area and initial developments, A few paragraphs ($\sim$350 words) on most recent and current developments.}   

%\section{Cloud formation modelling}\label{s:cloudm} 

The results of atmosphere models \cite{2013ApJ...775...80F,
  2017A&A...608A..70J,2009A&A...506.1367W,2009ApJ...699..564S,2013MNRAS.435.3159D,2014A&A...561A...1M,2015MNRAS.453.2412C,2016A&A...594A..48L,2016ApJ...829..115M},
or retrieval approaches
\cite{2011ApJ...729...41M,2017ApJ...834...50B,2017ApJ...848..127B,2020arXiv200612821M}
will depend on their individual components, their ability to model
physics and chemistry and to solve it numerically.  Cloud formation
modelling in Earth and solar system atmosphere modelling, e.g.
\cite{1978Icar...36....1R,2006ACP.....6.4175H}, has a long
tradition. It is essential that different modelling approaches are
followed in order to enable a benchmarking process. Cloud formation
modelling is also a key component in atmosphere modelling and
retrieval for exoplanets where the atmospheric gas is chemically and
dynamically very different to what is known from our solar system.
Cloud particles are tracers of the gas from which they form, through
which they move, which they deplete or enrich with respect to elements
abundances, and which they heat or cool due their opacity.  In order
to understand the chemical composition of exoplanet atmosphere, to
link this composition to their evolutionary state by interpreting
observations from CHEOPS, HST, JWST, Ariel, PLATO etc., cloud
formation needs to be understood in enough detail to be applicable to
the diversity of known exoplanets. Such a cloud model needs to be
applicable to the wide ranges of thermodynamics, radiation and
chemical regimes that exoplanets have presented us with. Here we lay
out the challenge of building such a cloud model, starting from our
kinetic modelling philosophy.  This is not a speedy exercise and
retrieval modeller may therefore still wish to hold on to representing
a cloud by a set of parameters like particle size, or by reducing
complexity by keeping physical consistency \cite{2019A&A...622A.121O}.

\subsection{Basic processes to form a cloud particle} 

\paragraph{Seed formation:} Unless an exoplanet has dust particles or other condensation seeds swapped into its atmosphere (e.g. condensing meteoritic dust, volcano outbreaks, sand storms), the formation of a cloud particles begins in the gas phase by chemical reactions of  those atoms/molecules/ions that are available in the atmosphere of a certain element composition. The element composition and the local thermodynamic conditions determine the first condensates. The formation of condensation seeds can be understood as a chain of chemical gas-gas reactions that progresses to molecules and clusters with increasing size and complexity. The challenge is to determine the required thermodynamic cluster data,  the free energy change associated with the formation of a cluster of size $N$, $\Delta G(N)$.  Assuming that no activation energy is required, one can formulate a rate network for all cluster sizes \cite{1998A&A...337..847P} or approximate $\Delta G(N)$ based on the concepts of classical nucleation theory \cite{2018A&A...614A.126L}. If cluster formation requires an activation energy, all individual chemical reaction will need consideration. Such fundamental concepts enable the modelling of the formation of mineral and organic seeds for a wide range of chemical situations if the required material data (e.g. $\Delta G(N)$) are available. Further chemical paths open up in highly irradiated planets where photochemistry leads to the formation of hydrocarbon molecules of increasing complexity after carbon is released by photochemically destroying  CO or CH$_4$  \cite{2018ApJ...853....7K,2014IJAsB..13..173R}. Two chemically independent ensembles of seed particles maybe expected in highly irradiated exoplanets, as shown for WASP-43b \cite{2020arXiv200514595H}. It needs to be understood, however, that a seed formation rate will depend on how much elements are consumed by further bulk growth.

\begin{figure}
    \centering
    \includegraphics[width=18cm]{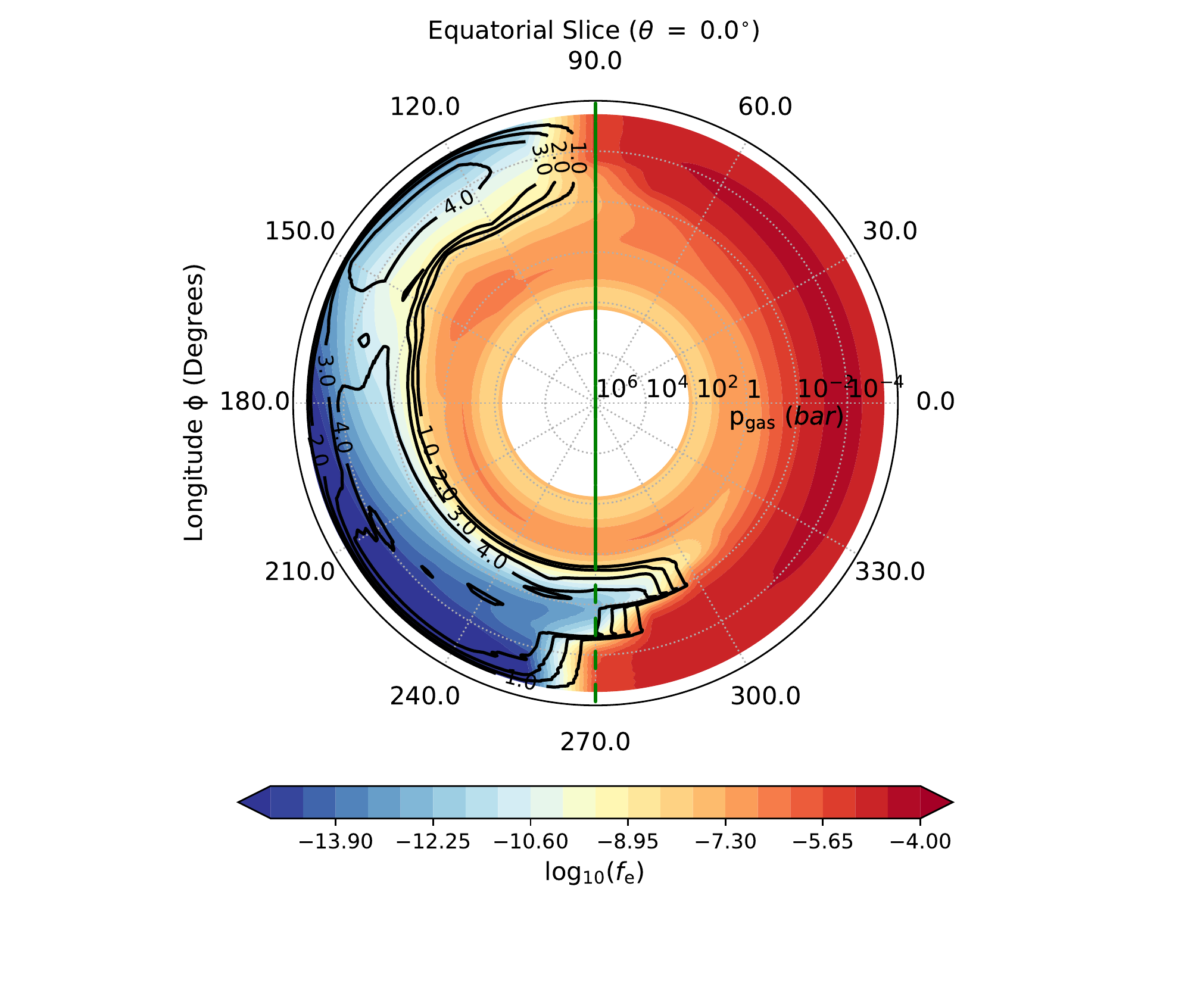}
    \caption{The exoplanet atmosphere challenge: Ultra-hot Jupiters combine an ultra-hot dayside forming an ionosphere with a high degree of ionisation (thermal degree of ionisation, $f_{\rm e}$ -- blue-yellow-red colour coded) with a cloudy nightside ($\rho_{\rm d}/\rho_{\rm gas}$ - black surface contour lines), shown here for HAT-P-7b as slice plot through the equatorial plane. The atmospheric pressure levels are shown as concentric rings [courtesy to D. Samra].}
    \label{fig:HAT-P-7b}

\end{figure}
\paragraph{Bulk growths:}
Once seed particles are present, the gas phase can easily condense on
these surfaces because non-photochemicaly-driven seed formation
requires a supercooling of the gas phase which is then highly
supersaturated. The kinetic bulk growth of cloud particles of mixed
composition by gas-surface reactions can efficiently be modelled
through a moment method that provides consistent information about
cloud particle composition, particles sizes and cloud extension
\cite{2006A&A...455..325H}.  A challenge are the details of the 100'ds
surface reactions which remain largely unexplored by laboratory
studied. A further process of increasing particles sizes is
coagulation, i.e. the collisional growth of existing cloud
particles. This can occur for hydrocarbon monomers or for mineral
cloud particles if the relative velocities are appropriate. Collisions
between cloud particles may also lead to shattering causing a decrease
of the mean particle size but an increase in particles number
\cite{2010A&A...513A..56G}. The arising challenge is that binning
methods work well for solving Smolochovsky's coagulation equation in a
mass conserving scheme but are challenged by particles of mixed
materials. Modelling the kinetic surface growth and the collisional
growth of existing dust particles furthermore requires to determine the cloud
particle velocity in potentially different frictional regimes.

\paragraph{Gravitational settling, frictional coupling, turbulent mixing and diffusion:}
Cloud particles move. Frictional coupling of cloud particles to the
gas phase determine how far cloud particles of certain sizes and
masses fall into the deeper atmospheric layers (and hence, keep
growing or begin to evaporate). They stand still (relative to the
surrounding gas) in the gas of complete friction (i.e. position)
coupling with the gas. Frictional coupling does also determine in how
far turbulent gas motion affects their further growth or even their
ionisation.  In case of complete coupling, turbulence will cause the
formation of intermittent cloud structures due to the emergence of
vorticity \cite{2004A&A...423..657H}. Diffusion is often used as
overarching term for transport against the atmospheric pressure
gradient and turbulent mixing as a driver of such diffusion. Per
definition, diffusion is driven by concentration gradients of
gas-species but also of cloud particles.  Each cloud particle size
would diffuse differently. Incorporating cloud particle diffusion in
cloud models has proven to be challenging, not the least because cloud
particle diffusion coefficients are unknown \cite{2020A&A...634A..23W}.

\paragraph{Neutral gas-phase chemistry:}
Modelling cloud formation requires knowledge about the local thermodynamic conditions, including the chemical composition for a set of element abundances. This, in turn, will depend on the presence of a condensed phase \cite{Oli2019}. The most complete set of information about  the gas-phase composition for almost all elements that we know from the Sun's atmosphere (incl. O, Mg, Fe, Si, Ca, Na, Al, Ti, $\ldots$) is provided by chemical equilibrium calculations (e.g. \cite{2018A&A...614A...1W}). Large gas-kinetic or photochemical calculations remain limited to molecules made of C/H/N/O and S and can, because of their complexity, be complete only unto a certain number of elements per molecule \cite{2016ApJS..224....9R}.  C/H/N/O are used to study the formation of hydrocarbon hazes and the triggering of possibly biology related molecules like HCN.   Such constrained chemical networks are very efficient for the study of specific processes, like for example,  lightning chemistry \cite{2016ApJS..224....9R} and biomolecules on the surface of charged cloud particles \cite{2014IJAsB..13..165S}. The challenge is far more fundamental, namely, to build kinetic rate networks that link to the need of the nucleation modelling outlines above and to model spectacular events like lightning in the exoplanet context.

\subsection{Ionisation process}
Atmospheres of exoplanets can be highly irradiated such that the
atmospheric gas is photoionised and/or ionised thermally. The
ionisation efficiency may vary from day to night side leading to a
highly ionised and cloud-free dayside and a cloudy nightside with just
an ionosphere ontop on hot rocky planets like 55 Cnc e \cite{2017MNRAS.472..447M}  as on super-hot Jupiters \cite{2019A&A...626A.133H}. The photoionisation can reach the upper cloud
layers, but a more realistic process is that currents from the ionised
atmosphere form and transport charges onto the surface of the cloud
particles, similar to Earth.

Figure~\ref{fig:HAT-P-7b} demonstrate that for the ultra-hot Jupiter HAT-P7b the day/night difference of the thermal degree of ionisation (colour coded). An ionosphere is extending deep into the dayside atmosphere but remains rather shallow on the nightside. The nightside, in contrast is covered in clouds (solid black lines show the cloud-particle mass loss in term of the dust-to-gas mass density ratio) which also assymetrically cover the two terminator regions.

%\section{Important questions/goals} %500-600 words 
%\textcolor{red}{A summary of the major questions/goals to be addressed in this particular area.} 

\section{Opportunities} % 200-300 words 
%\textcolor{red}{A summary of the key opportunities in this area} 

Understanding if the atmosphere of an exoplanet enables biological
activity requires us to understand the interaction of every-day
processes that occur on Earth in a far larger parameter range than
available on Earth. This is why laboratory experiments on Earth are
unfortunately only of limited value for exoplanet research in its whole
diversity of exoplanet specisims. The exoplanet parameter range is
increased by being linked to the evolutionary state of the planet, and
its formation history. We therefore require complex models to study in
detail the physico-chemistry of exoplanet atmospheres, what is more,
we require several of those models to be developed in parallel to be
able to follow the fundamental research ethics of testing and
comparing results by different means and apertures. The use of only
one model defeats the purpose of research and the necessary
falsification of results.

\smallskip
Building virtual laboratories by virtue of complex models is the only
path to understand the vast exoplanet diversity in depth, and to
decode observations that can only be taken for selected times (or time
intervals) and for selected wavelengths (or wavelength bands) for
selected objects.  Monitoring exoplanets of different kinds, that
orbit different stellar types, in wavelengths and in time is
unfeasible in the foreseeable future. The purpose of such monitoring
could be as far-fetched as finding a proper landing side or as
fundamental as understanding the different planets as global
objects. The EAS Voyage 2050 white papers on exoplanet research remain
focused by necessity, including spectropolarimetry
\cite{2019BAAS...51c..33F}, spatial resolved mid-IR of thermal
emission  \cite{2019arXiv190801316Q} and high-contrast imaging
\cite{2019arXiv190801803S} missions as next steps beyond JWST, Ariel and the
ELT-class telescopes.  Modelling the underpinning exoplanet atmosphere
physics and chemistry will be essential.  Virtual laboratories already
enables us to explore exoplanet atmosphere with respect to specific
processes (like the effect of cloud formation), or as the sum of all
known processes (like the emerging of weather pattern), and in
particular beyond the singled-out terrestrial conditions (e.g. for a
diversity of element abundances).

\section{Challenges}  % 200-300 words
%\textcolor{red}{A summary of the main challenges in this area} 

{\sf Outstanding Challenges} to enable virtual laboratories to meet the observational challenge include:\\[0.2cm]
-- the element abundances that determine the planet's atmosphere chemistry\\
-- chemical and cluster data for all possible nucleation species \\
-- thermochemical data for reaction rates for element other than C/N/O/H/(S) \\
-- thermochemical data for gas-surface reactions, incl. sticking probabilities\\
-- opacity data for condensates across large wavelengths ranges and crystalline species\\
-- opacity modelling for non-spherical and charged cloud particles.

\smallskip\noindent
Meeting these challenges will enable us to decipher the evolutionary state and the planet's formation mechanisms  spectroscopically, and to determine with confidence the state of habitability of an extrasolar planet.

%---------------- Including references and figures ------
%\textcolor{red}{Citations can be made as follows \citep{fisher2014}. A template is included for your bib file with filename ``references\_code.bib", where code is a unique code, e.g. initials of your essay title. It is recommended that any figure included should aim to summarize the area or be a representation of a central aspect of the area. Figures can be included anywhere in the text as follows.}

%\begin{figure}[t]
%\begin{center}
%\includegraphics[width=0.25\textwidth]{fig1.eps}
%\end{center}
%\caption{This is figure 1.}
%\end{figure}

%\newpage
\bibliographystyle{plain}
\bibliography{references_clouds}
\end{document}